\documentstyle[times,epsfig]{jaa}

\begin{document}
\title[Spectral Variation of PMN J0948+0022]{Spectral Variation of NLS1 Galaxy PMN J0948+0022}
\author[X. N. Sun et al.]
        {X. N. Sun $^{1,3}$, Jin Zhang $^{2,3}$\thanks{e-mail:zhang.jin@hotmail.com}, Y. Lu $^{2}$, E. W. Liang $^{1,2,3}$, S. N. Zhang$^{2,4}$ \\
$^1$ Department of Physics and GXU-NAOC Center for Astrophysics and Space Sciences, \\
  Guangxi University, Nanning, 530004, China\\
$^2$ National Astronomical Observatories, Chinese Academy of Sciences, Beijing, 100012, China\\
$^3$ Key Laboratory for the Structure and Evolution of Celestial Objects,\\
  Chinese Academy of Sciences, Kunming, 650011, China\\
$^4$ Key Laboratory of Particle Astrophysics, Institute of High Energy Physics,\\
  Chinese Academy of Sciences, Beijing, 100049, China}

\maketitle \label{firstpage}
\begin{abstract}
Four well-sampled spectral energy distributions (SEDs) of PMN J0948+0022 are fitted with the syn+SSC+EC model to
derive the physical parameters of its jets and to investigate the spectral variations of its SEDs. A tentative
correlation between the peak luminosity ($L_{\rm c}$) and peak frequency ($\nu_{\rm c}$) of its inverse Compton
(IC) bump is found in both the observer and co-moving frames, indicating that the variations of luminosity are
accompanied with the spectral shift. A correlation between $L_{\rm c}$ and $\delta$ is found, and thus the
magnification of the external photon field by the bulk motion of the radiation regions is an essential reason
for the spectral variation since the IC bump of PMN J0948+0022 is dominated by the EC process.
\end{abstract}

\section{Introduction}
\label{sec:intro} Narrow-Line Seyfert 1 (NLS1) Galaxies are a relatively peculiar subclass of active galactic
nuclei (AGNs), which are characterized by their optical spectra with narrow permitted lines (FWHM (H$\beta$) $<$
2000 km s$^{-1}$), the ratio of [O III]$\lambda$5007$/$H$\beta$ $<$ 3, and the bump of Fe II (e.g., Pogge 2000).
NLS1s are also interesting for their low masses of central black holes and high accretion rate. NLS1s are
generally radio quiet; only a small percentage of them is radio-loud ($<7\%$; Komossa et al. 2006). So far, four
NLS1s are confirmed to be GeV emission sources by Fermi/LAT, which should have relativistic jets as predicted by
Yuan et al. (2008).

PMN J0948+0022 (z = 0.5846) is the first NLS1 detected in GeV band by Fermi/LAT. The inverted spectrum in radio
band indicates the presence of a relativistic jet viewed at small angles (Zhou et al. 2003), which is similar to
the properties of blazars. The significant variabilities of PMN J0948+0022 are detected, especially in GeV band
(Abdo et al. 2009; Foschini et al. 2012). It is found that the observed luminosity variations are usually
accompanied with the shift of peak frequencies of the SEDs, similar to some GeV-TeV BL Lacs (Zhang et al. 2012).
The abundant broadband observational data provide an opportunity to study the physical mechanism of the spectral
variations of PMN J0948+0022.

\begin{figure}
\centering
\includegraphics[angle=0,scale=0.35]{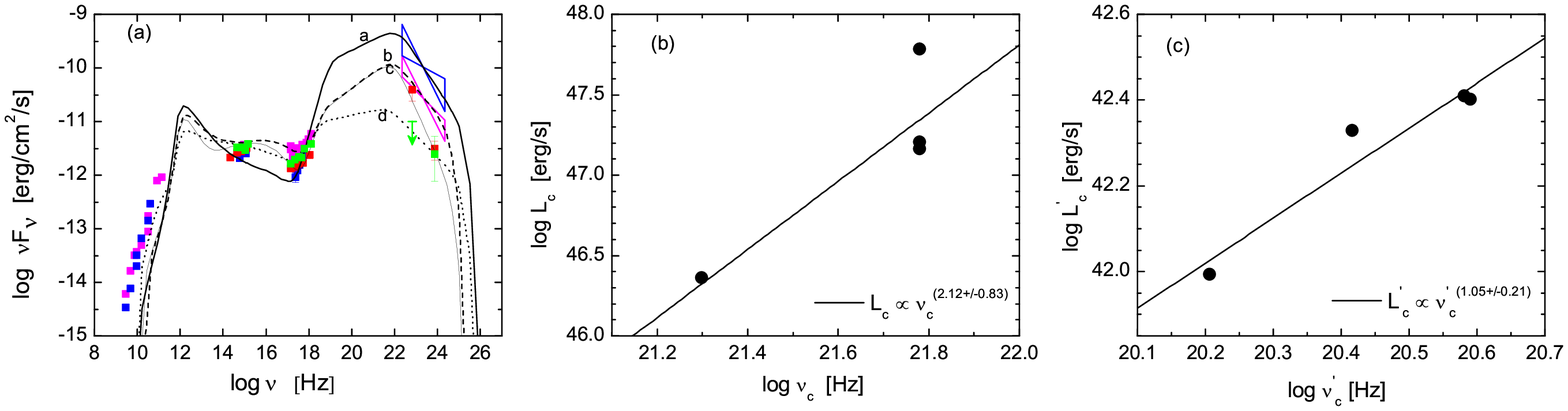}
\caption{{\em Panel a}---Observed SEDs ({\em scattered data points}) with model fitting ({\em lines}) for PMN
J0948+0022. The four SEDs are marked as states ``a" ({\em blue points and thick solid line}), ``b" ({\em magenta
points and dashed line}), ``c" ({\em red points and thin solid line}), and ``d" ({\em green points and dotted
line}), respectively. {\em Panels b, c}---$L_{\rm c}$ as a function of $\nu_{\rm c}$ in both the observer ({\em
Panel b}) and co-moving ({\em Panel c}) frames.}
\end{figure}

\section{SED Selection and Modelling }
\label{sec:using} We compile the observed broadband SEDs of PMN J0948+0022 from literature. Four available SEDs
as shown in Figure 1(a), are defined as SEDs ``a", ``b", ``c", and ``d" according to their peak luminosity of the IC
bump of the SEDs, respectively. The data of SEDs ``a" and ``b" are from Foschini et al. (2012) and are obtained
with the observations in 2010 July 8th and 2011 October 9th-12th, respectively. The SEDs ``c" and ``d" of this
source are taken from the observations in 2009 June 14th and 2009 May 5th (Abdo et al. 2009).

The broadband SEDs of PMN J0948+0022 are similar to the typical FSRQs and thus the $\gamma$-ray emission should
be dominated by jet emission. We use the syn+SSC+EC model to fit its SEDs because the contributions of the
external field photons from the broad line region (BLR) need to be considered. The total luminosity of the BLR
is calculated using the luminosity of its emission lines (Zhou et al. 2003) with equation (1) given in Celotti
et al. (1997). The size of the BLR is calculated using the BLR luminosity with equation (23) in Liu \& Bai
(2006). The energy density of the BLR measured in the co-moving frame is $U^{'}_{\rm
BLR}=6.76\times10^{-3}\Gamma^{2}$ erg cm$^{-3}$, where we take $\Gamma\sim\delta$. The minimum variability
timescale is taken as $\Delta t=12$ hr. The details of the model and the strategy for constraining its
parameters constraints can be found in Zhang et al. (2012).

\begin{figure}
\centering
\includegraphics[angle=0,scale=0.35]{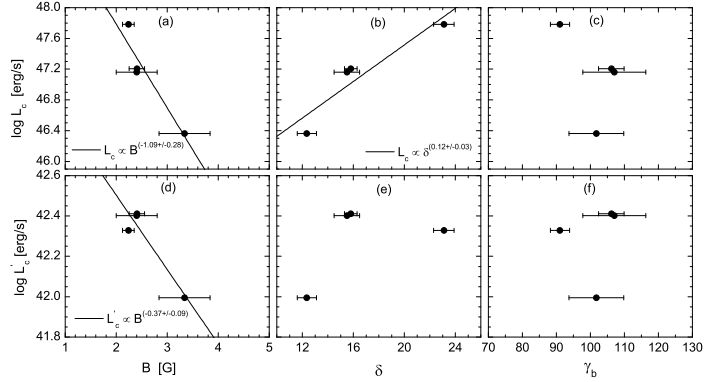} \caption{$L_{\rm c}$ as functions of $B$, $\delta$, and $\gamma_{\rm b}$ in both the observer ({\em
three top panels}) and co-moving ({\em three bottom panels}) frames.}
\end{figure}

The SEDs of PMN J0948+0022 are well explained with the syn+SSC+EC model, as shown in Figure 1(a). The EC component of the SEDs for PMN J0948+0022 presents a
further constraint on $\delta$, and thus makes a tighter constraint on $\delta$ and $B$ than that for BL Lacs in
Zhang et al. (2012). The fitting parameters of SEDs for PMN J0948+0022 are also more tightly clustered; the
magnetic field strength $B$ is from $2.24\pm0.11$ G to $3.34\pm0.50$ G, the Doppler factor is from $12.4\pm0.8$
to $23.0\pm0.8$, and the break Lorenz factor of electrons is from $91\pm3$ to $107\pm9$.

\section{Spectral Variation of IC Bump}
\label{sec:frontmatter}

The broadband SEDs of PMN J0948+0022 are dominated by the EC process, and there are significant variabilities in
GeV band. The peak luminosity ($L_{\rm c}$) as a function of peak frequency ($\nu_{\rm c}$) of the IC bump in
both the observer and co-moving frames are given in Figures 1(b),(c). A tentative correlation between the peak
luminosity and peak frequency in both the observer and co-moving frames is found, i,e., $L_{\rm c} \propto
\nu_{\rm c}^{(2.12\pm0.83)}$ with $r = 0.87$ (Pearson correlation coefficient) and $p=0.13$ (chance
probability), and $L^{'}_{\rm c}\propto \nu^{'(1.05\pm0.21)}_{\rm c}$ with $r = 0.96$ and $p=0.04$,
respectively, indicating that the luminosity variations of the IC bump are accompanied with a spectral shift.

To investigate the possible physical reason of this phenomenon, we show the IC peak luminosity as functions of
$B$, $\delta$, and $\gamma_{\rm b}$ in both the observer and co-moving frames in Figure 2. It can be found that:
(1) Both $L_{\rm c}$ and $L^{'}_{\rm c}$ are anti-correlated with $B$. The Pearson correlation analysis and the
best liner fits yield $L_{\rm c} \propto B^{(-1.09\pm0.28)}$ with $r=-0.94$, $p=0.06$ and $L^{'}_{\rm c}\propto
B^{(-0.37\pm0.09)}$ with $r=-0.94$, $p=0.06$. (2) $L_{\rm c}$ seems to be correlated with $\delta$ with $r=0.93$
and $p=0.07$. (3) No correlation between $\gamma_{\rm b}$ with $L_{\rm c}$ and $L^{'}_{\rm c}$ is found, which
may be due to the uncertainties of the synchrotron radiation peak for the SEDs. These facts indicate that the
spectral variations of the IC peak for PMN J0948+0022 may be attributed to the variations of $\delta$ and $B$, similar to the results of a typical FSRQ 3C 279 (Zhang et al. 2013).

The significant variations of the IC peak for PMN J0948+0022 in GeV band are dominated by
the EC process. The energy density of the external photon field would be magnified by $\Gamma^2$ and the energy
of the seed photons would be magnified by $\Gamma$ due to the motion of the emitting regions, hence a small
variation of $\delta$ would result in significant variations of $\nu_{\rm c}$ and $L_{\rm c}$. As mentioned
above, $B$ is also anti-correlated with $L_{\rm c}$ in both the observer and co-moving frames, indicating the
variations of $B$ for this source between different states, which are also accompanied with the variations of
$\delta$, might be linked to the variations of some intrinsic physical parameters of the center black hole, such
as the disk accretion rate or the corona (Zhang et al. 2013). The instabilities of corona or disk accretion rate may
result in the variations of the jet physical condition and the variations of jets emission.

\section{Conclusion}
The SEDs observed at four epochs for PMN J0948+0022, which can be explained well with the syn+SSC+EC model, are
compiled from literature to investigate its spectral variation. A tentative correlation between the peak
luminosity and peak frequency of its IC bumps is found, indicating that a higher GeV luminosity corresponds to a
harder spectrum for the emission in the GeV band, similar to the properties of some blazars. The SEDs of PMN
J0948+0022 are dominated by the EC bumps and thus the magnification of the external photon field by the bulk
motion of the radiation regions is an essential reason for the spectral variation. The variations of $B$ and
$\delta$ for PMN J0948+0022 between different states may be produced by the instabilities of the corona or the
disk accretion rate.

\section{Acknowledgments}
This work is supported by the National Basic Research Program (973 Programme) of China (Grant 2009CB824800), the
National Natural Science Foundation of China (Grants 11078008, 11025313, 11133002, 10725313), Guangxi Science Foundation (2011GXNSFB018063, 2010GXNSFC013011), and Key Laboratory for the Structure and Evolution of Celestial Objects of Chinese Academy of Sciences.

\end{document}